\documentclass[
  amsfonts,
  amsmath,
  amssymb,
  pra,
  twocolumn,
  superscriptaddress,
]{revtex4-2}

\usepackage{appendix}[titletoc]
\usepackage[english]{babel}
\usepackage[justification=centering]{caption}
\usepackage{float}
\usepackage{graphicx}
\usepackage{hyperref}
\usepackage[utf8]{inputenc}
\usepackage{layouts}
\usepackage{multirow}
\usepackage{optidef}
\usepackage{physics}
\usepackage{setspace}
\usepackage{diagbox}
\usepackage[labelformat=simple]{subcaption}
\usepackage{xcolor}

\definecolor{darkgreen}{RGB}{46, 184, 46}

\restylefloat{table}

\addto\captionsenglish{%
}

\begin{document}

\title{Robust Quantum Optimal Control with Trajectory Optimization}

\author{Thomas Propson}
\email{tcpropson@uchicago.edu}
\affiliation{
  James Franck Institute, University of Chicago, Chicago, Illinois 60637, USA
}
\affiliation{
  Department of Physics, University of Chicago, Chicago, Illinois 60637, USA
}
\author{Brian E. Jackson}
\affiliation{
  Robotics Institute, Carnegie Mellon University, Pittsburgh, Pennsylvania 15213, USA
}

\author{Jens Koch}
\address{Department of Physics and Astronomy, Northwestern University, Evanston, Illinois 60208, USA}
\address{Northwestern–Fermilab Center for Applied Physics and Superconducting Technologies, Northwestern University, Evanston, Illinois 60208, USA}

\author{Zachary Manchester}
\affiliation{
  Robotics Institute, Carnegie Mellon University, Pittsburgh, Pennsylvania 15213, USA
}
\author{David I. Schuster}
\affiliation{
  James Franck Institute, University of Chicago, Chicago, Illinois 60637, USA
}
\affiliation{
  Department of Physics, University of Chicago, Chicago, Illinois 60637, USA
}
\affiliation{
  Pritzker School of Molecular Engineering, University of Chicago, Chicago, Illinois 60637, USA
}

\date{\today}

\begin{abstract}
  The ability to engineer high-fidelity gates on quantum processors in the presence of
  systematic errors remains the primary barrier to achieving quantum advantage.
  Quantum optimal control methods have proven effective in experimentally
  realizing high-fidelity gates, but they require exquisite calibration to be performant.
  We apply robust trajectory optimization techniques to suppress gate errors arising from system
  parameter uncertainty.
  We propose a derivative-based approach that maintains
  computational efficiency by using forward-mode differentiation.
  Additionally, the effect of depolarization on a gate is typically modeled by
  integrating the Lindblad master equation,
  which is computationally expensive.
  We employ a computationally efficient model
  and utilize time-optimal control to achieve high-fidelity gates in the presence of depolarization.
  We apply these techniques to a fluxonium qubit and suppress
  simulated gate errors due to parameter uncertainty below $10^{-7}$ for
  static parameter deviations on the order of $1\%$.
\end{abstract}

\maketitle

\section{Introduction}
Quantum optimal control (QOC) is a class of optimization
algorithms for accurately and efficiently manipulating quantum systems.
Early techniques were proposed for nuclear magnetic resonance experiments
\cite{vandersypen2005nmr, kehlet2004improving, khaneja2005optimal,
  maximov2008optimal, nielsen2010optimal, skinner2003application, tosner2009optimal},
and applications now include superconducting circuits \cite{abdelhafez2020universal,
  chakram2020multimode, egger2013optimized, fisher2010optimal, gokhale2019partial,
  huang2014optimal, heeres2017implementing, kelly2014optimal, leng2019robust,
  leung2017speedup, li2020fast,
  liebermann2016optimal, reinhold2019controlling,
  rebentrost2009optimal, rebentrost2009optimal2, spiteri2018quantum,
  sporl2007optimal},
neutral atoms and ions \cite{brouzos2015quantum,
  de2008optimal, grace2007optimal, goerz2011quantum, guo2019high, jensen2019time,
  larrouy2020fast, nebendahl2009optimal, omran2019generation,
  rosi2013fast,
  treutlein2006microwave, van2016optimal},
nitrogen-vacancy centers in diamond \cite{chou2015optimal,
  dolde2014high, geng2016experimental,
  nobauer2015smooth, poggiali2018optimal, rembold2020introduction, tian2019optimal},
and Bose-Einstein condensates \cite{amri2019optimal, doria2011optimal,
  sorensen2019qengine, sorensen2018quantum}.
In the context of quantum computation,
optimal control is employed to achieve high-fidelity gates
while adhering to experimental constraints.
Experimental errors such as parameter drift, noise, and
finite control resolution cause the system to deviate
from the model used in optimization, hampering
experimental performance
\cite{chakram2020multimode, heeres2017implementing, klimov2020snake,
  omran2019generation, reinhold2019controlling}.
Robust control improves upon
standard optimal control by encoding
model parameter uncertainties
in optimization objectives, yielding performance
guarantees over a range of parameter values \cite{Zhou97,Morimoto00,Manchester18}.
We adapt robust control techniques from the robotics community to mitigate
parameter-uncertainty errors for
a superconducting fluxonium qubit.

Analytically-derived control pulses that mitigate parameter-uncertainty
errors include composite pulses \cite{cummins2000use, cummins2003tackling,
  kupce1995stretched, merrill2014progress},
pulses designed by considering dynamic and geometric phases
\cite{han2020experimental, xu2020nonadiabatic}, and
pulses obtained with the DRAG scheme \cite{motzoi2009simple}.
As compared to analytical techniques, QOC is advantageous for
designing pulses that consider all experimental constraints and
performance tradeoffs
\cite{leung2017speedup},
and for constructing operations without a known analytic solution
\cite{chakram2020multimode, heeres2017implementing}.
Accordingly, recent work has sought to achieve robustness in QOC
frameworks using closed-loop methods
\cite{egger2014adaptive, feng2018gradient,
  li2017hybrid, wittler2020integrated}
and open-loop methods
\cite{ball2020software, carvalho2020error, allen2019robust, reinhold2019controlling,
  rembold2020introduction, kosut2013robust, niu2019universal, khaneja2005optimal}.

In this work, we study three open-loop robust control techniques that
make the quantum state trajectory less sensitive to the uncertainties
of static and time-dependent parameters:
\begin{enumerate}
\item A sampling method, similar to the work in Refs.~
  \cite{allen2019robust, khaneja2005optimal, reinhold2019controlling, rembold2020introduction}.
\item An unscented sampling method \cite{howell2020direct, lee2013sigma, thangavel2020robust}
  adapted from the unscented transform \cite{julier2004unscented,
    uhlmann1995dynamic} used in state estimation.
  \item A derivative method, which penalizes the sensitivity of the quantum state trajectory
    to uncertain parameters.
\end{enumerate}
We apply these techniques to the fluxonium qubit presented in \cite{zhang2020universal}.
We also show that QOC can solve important problems associated with
fluxonium-based qubits: exploiting the dependence of $T_{1}$ on the controls
to mitigate depolarization
and synchronizing the phase of qubits with distinct frequencies.
To ameliorate depolarization,
we perform time-optimal control and
employ an efficient depolarization model
for which the computational cost is independent of the
Hilbert space dimension.
Leveraging recent advances in trajectory optimization within the field of robotics, we
solve these optimization problems using ALTRO (Augmented Lagrangian TRajectory Optimizer)
\cite{howell2019altro}, which can enforce constraints on
the control fields and the quantum state trajectory.

This paper is organized as follows.
First, we describe ALTRO in the context of QOC
in Sec. \ref{sec:background}.
We outline realistic constraints for operating the fluxonium and
define the associated QOC problem in Sec. \ref{sec:fluxonium}.
Then, we formulate a method for suppressing depolarization
in Sec. \ref{sec:longitude}. Next, we describe three techniques for achieving
robustness to static parameter uncertainties in Sec. \ref{sec:static}. We
adapt the same techniques to mitigate 1/$f$ flux noise
in Sec. \ref{sec:stochastic}.

\section{Background \label{sec:background}}
In this section, we
review the QOC problem statement
and describe the ALTRO solver \cite{howell2019altro}.
QOC concerns a vector $\mathbf{a}(t)$ of time-dependent control fields
that steer the evolution of a quantum state $\ket{\psi(t)}$.
The evolution of the state is governed by the time-dependent
Schr{\"o}dinger equation (TDSE),
\begin{equation}
  i \hbar \frac{d}{dt} \ket{\psi(t)} = H(\mathbf{a}(t), t) \ket{\psi(t)}.
  \label{eq:tdse}
\end{equation}
The Hamiltonian $H(\mathbf{a}(t), t)$ is determined by the quantum system and the external control fields.
The QOC problem is to find the
controls that minimize a functional $J[\mathbf{a}(t)]$,
which we call the objective.
To make the problem numerically tractable,
the quantum state and controls are discretized into $N$ time steps,
$\ket{\psi(t_{k})} \to \ket{\psi_k}$ and $\mathbf{a}(t_{k}) \to \mathbf{a}_k$ where
$t_{k} = t_{k - 1} + \Delta t$ and $k \in \{1, ..., N\}$.
In the case of a single state-transfer problem, the objective is
the infidelity of the time-evolved final state $\ket{\psi_N}$ and
the intended target state $\ket{\psi_T}$,
$J(\mathbf{a}) = 1 - {\lvert \braket{\psi_{T}}{\psi_{N}(\mathbf{a})} \rvert}^{2}$.
Standard QOC solvers compute derivatives of the objective $\nabla J(\mathbf{a})$,
which can easily be used to implement first-order optimization methods
\cite{machnes2015tunable, khaneja2005optimal, leung2017speedup, goerz2019krotov}.

Alternatively, the QOC problem can be formulated as a trajectory optimization problem 
and solved using specialized solvers developed by the robotics community
\cite{Schulman13, Tedrake16, Hereid2017FROST, howell2019altro}.
The objective $J(\mathbf{a})=\sum_k \ell_{k}(\mathbf{x}_{k}, \mathbf{u}_{k})$
is expressed in terms of the
cost function at each time step $\ell_k$, where
$\mathbf{x}_{k}$ is the augmented state vector
and $\mathbf{u}_{k}$ is the augmented control vector.
We use the term \emph{augmented} because these
vectors contain all of the relevant variables in the optimization problem,
not just the quantum state and the control fields,
for an example see Sec. \ref{sec:fluxonium}.
The augmented control contains all variables that the experimentalist
may manipulate, and the augmented state contains all variables that depend
on those in the augmented control.
The variables in the augmented states depend on those in the augmented
controls as defined by the differential equations governing the physical system, which are
encoded in the discrete relation
$\mathbf{x}_{k + 1} = \mathbf{f}(\mathbf{x}_{k}, \mathbf{u}_{k})$.
For QOC, $\mathbf{f}(\mathbf{x}_{k}, \mathbf{u}_{k})$ -- which we call
the discrete dynamics function -- propagates the quantum state by
integrating the TDSE \eqref{eq:tdse}
using a Runge-Kutta method \cite{jorgensen2011numerical}
or an exponential integrator \cite{auer2018magnus, berland2006solving, einkemmer2017performance,
  shillito2020fast}.

We incorporate constraints on the augmented controls and states
by formulating them as inequalities  $\mathbf{g}_{k}(\mathbf{x}_{k}, \mathbf{u}_{k}) \leq \mathbf{0}$
or equalities $\mathbf{h}_{k}(\mathbf{x}_{k}, \mathbf{u}_{k}) = \mathbf{0}$.
The constraint functions $\mathbf{g}_k$ and $\mathbf{h}_k$ may be vector-valued to encode multiple constraints,
and equalities and inequalities are understood component-wise.
To quantify constraint satisfaction, we define each constraint's \emph{violation}
as the magnitude of its deviation:
$\textrm{max}(g(\cdot), 0)$ or $\abs{h(\cdot)}$, where $g$ and $h$
are components of constraint functions $\mathbf{g}_{k}$ and $\mathbf{h}_{k}$,
respectively.
Stated concisely, the trajectory optimization problem is:
\begin{mini!}[2]
  {\substack{\mathbf{x}_1, \ldots, \mathbf{x}_N \\\mathbf{u}_1, \ldots, \mathbf{u}_{N - 1}}}{
    \sum^{N}_{k = 1}\ell_{k}
    (\mathbf{x}_k,\mathbf{u}_k),}{}{} \label{eq:gcostfun}
  \addConstraint{\mathbf{x}_{k+1} = \mathbf{f}(\mathbf{x}_k,\mathbf{u}_k) \ \forall \ k}
  \label{eq:gdyncon}
  \addConstraint{\mathbf{g}_k(\mathbf{x}_k,\mathbf{u}_k) \leq \mathbf{0} \ \forall \ k}
  \label{eq:ineqcon}
  \addConstraint{\mathbf{h}_k(\mathbf{x}_k,\mathbf{u}_k) = \mathbf{0} \ \forall \ k.}
  \label{eq:eqcon}
\end{mini!}
We have formulated the problem such that the cost and constraint functions at
time step $k$ may only depend on the augmented control and state at time step $k$.
Although this structure may appear limiting, the problem can
typically be reformulated to accomodate any cost or constraint function,
for an example see Sec. \ref{sec:fluxonium}, and the ALTRO solver,
which we introduce in the following discussion, exploits this
structure to efficiently solve the problem.

Standard techniques for solving \eqref{eq:gcostfun}-\eqref{eq:eqcon} typically
fall into two categories: direct methods \cite{Hargraves87, kelly2017introduction}
and indirect methods \cite{betts1998survey}. For indirect methods,
the augmented controls are the \emph{decision variables}, i.e., the
variables the optimizer adjusts to solve the problem.
The augmented states are obtained from the augmented controls using the discrete dynamics function,
and they are used to evaluate derivatives of the cost functions.
Then, the derivative information is employed to update the augmented controls.
This approach is taken by standard QOC solvers such as GOAT \cite{machnes2015tunable},
GRAPE \cite{khaneja2005optimal, leung2017speedup}, and Krotov's method \cite{goerz2019krotov}.
Conversely, direct methods treat both the augmented controls and states as decision
variables. In addition to minimizing the cost functions, the optimizer uses derivative information
for the discrete dynamics function to satisfy the dynamics constraint
\eqref{eq:gdyncon} to a specified tolerance.
In this sense, the TDSE \eqref{eq:tdse} is a constraint that may be violated
for intermediate steps of the optimization, where the quantum states need not be physical.
The direct approach lends itself to a nonlinear program formulation, for which
a variety of general-purpose solvers exist \cite{gill2005snopt, wachter2006implementation}.

Recent state-of-the-art solvers, such as ALTRO,
combine the indirect and direct methods in a two-stage approach.
First, ALTRO employs an indirect solving stage using
the iterative linear-quadratic regulator (iLQR) algorithm
\cite{Li2004a} as the internal solver of an augmented Lagrangian method (ALM)
\cite{lantoine2012hybrid, plancher2017constrained, nocedal2006numerical}.
In the second direct stage, ALTRO uses a projected Newton method
\cite{bertsekas1982projected, rao1998application}.
Next, we provide a more detailed summary of these two stages.

iLQR is an indirect method
for minimizing the objective subject to the dynamics constraint,
i.e., solving \eqref{eq:gcostfun}-\eqref{eq:gdyncon}.
First, iLQR uses an initial guess for the augmented controls to obtain the 
augmented states with the discrete dynamics function.
iLQR then constructs quadratic models for each cost function using
their zeroth-, first- and second-order derivatives in a Taylor expansion
about the current augmented controls and states.
These models are used with a recurrence relation between time steps
to obtain the locally optimal update for the augmented controls.
This recurrence relation is possible to derive in closed
form because cost function contributions come only from the augmented
control and state at a single time step \cite{mayne1966a}.
Finally, a line search \cite{zhang2006global}
is performed in the direction of the locally optimal update to ensure a
decrease in the objective.
This procedure is repeated until convergence is reached.

While indirect solvers like iLQR are computationally
efficient and maintain high accuracy for the discrete dynamics
throughout the optimization, they cannot handle
nonlinear equality and inequality
constraints \eqref{eq:ineqcon}-\eqref{eq:eqcon}.
For QOC, a popular approach to handle such constraints
is to add the constraint functions to the objective
\cite{heeres2017implementing, leung2017speedup, reinhold2019controlling, niu2019universal}.
However, this strategy does not guarantee that the constraints
are satisfied as the solver trades
minimization of the cost functions and constraint functions against each other.
ALM remedies this issue by adaptively adjusting a Lagrange multiplier estimate
for each constraint function to ensure the constraints are satisfied.
ALM adds terms that are linear and quadratic in the constraint functions
to the objective. Then, the new objective is minimized with
iLQR. If the solution obtained with iLQR does not satisfy the constraints,
the prefactors for the constraint terms in the objective are increased
intelligently and the procedure is repeated.

ALM converges superlinearly, but poor numerical conditioning may lead
to small decreases in the constraint violations near the locally optimal solution
\cite{bertsekas2014constrained}.
To address this shortcoming, ALTRO
projects the solution from the ALM stage onto the constraint manifold using
a (direct) projected Newton method, achieving ultra-low
constraint violations $\sim 10^{-8}$.
For more information on the details of the ALTRO
solver, see Refs. \cite{howell2019altro, Jackson2020altroc}.

As opposed to standard QOC solvers, ALTRO
can satisfy constraints on both the control fields and quantum states to tight tolerances.
This advantage is crucial for this work, where multiple medium-priority cost functions
are minimized subject to many high-priority constraints.

\section{QOC for the Fluxonium \label{sec:fluxonium}}
In the following, we optimize quantum gates
for the superconducting fluxonium qubit -- a promising
building block for quantum computers due to its high
coherence times
\cite{earnest2018realization, lin2018demonstration,
  manucharyan2009fluxonium, somoroff2021millisecond, nguyen2019high,
  zhang2020universal}.
In this section, we use the trajectory optimization
formalism \eqref{eq:gcostfun}-\eqref{eq:eqcon}
to define the optimization problem \eqref{eq:costfun}-\eqref{eq:amp_con},
which we extend in subsequent sections to account
for experimental error channels.
To high accuracy, we approximate the fluxonium Hamiltonian near the flux-frustration
point as a two-level system:
\begin{align}
  H/h &= f_{q} \frac{\sigma_{z}}{2} + a(t) \frac{\sigma_{x}}{2}.
  \label{eq:hamiltonian}
\end{align}
Here, $f_{q}$ is the qubit frequency at the flux-frustration point,
$a(t)$ is the control governing the flux offset from the flux-frustration point,
$h$ is Planck's constant, and $\sigma_{z}, \sigma_{x}$
are Pauli matrices. Although the coherent dynamics can be described with this two-level
system model, our noise model, experimental constraints, and system parameters
consider the full system, and they are representative of the fluxonium
presented in \cite{zhang2020universal}.

First, we introduce the augmented control and state for the fluxonium gate problem.
Since the ALTRO implementation we use does not currently
support complex numbers, we represent the quantum states
in the isomorphism $\mathcal{H}(\mathbb{C}^{n})
\cong \mathcal{H}(\mathbb{R}^{2n})$ given in \cite{leung2017speedup},
\begin{equation}
  H \ket{\psi} \cong \begin{pmatrix} H_{\textrm{re}} & -H_{\textrm{im}}
    \\ H_{\textrm{im}} & H_{\textrm{re}}\end{pmatrix}
  \begin{pmatrix} \ket{\psi}_{\textrm{re}} \\ \ket{\psi}_{\textrm{im}}\end{pmatrix}.
  \label{eq:isomorphism}
\end{equation}
We use $\psi$ -- abandoning bra-ket notation -- to denote the real representation of a state
given by the right-hand-side of \eqref{eq:isomorphism}.
To refer to the discrete moments of the flux, we introduce the notation
$\int_{t} a_{k} \equiv \int^{t_{k}}_{t_{1}} a(t) \ \mathrm{d}t$,
$a_{k} \equiv a(t_{k})$,
$\mathrm{d}^{n}_{t} a_{k} \equiv \mathrm{d}^{n}a(t)/{\mathrm{d}t}^{n} \lvert_{t = t_{k}}$.
The augmented control and state are:
\begin{equation}
  \begingroup
  \renewcommand*{\arraystretch}{1.3}
  \mathbf{u}_{k} = \begin{pmatrix} \mathrm{d}^{2}_{t} a_{k} \end{pmatrix}, \quad
  \mathbf{x}_{k} = \begin{pmatrix} \psi^{0}_{k} \\ \psi^{1}_{k}
    \\ \int_{t} a_{k} \\ a_{k} \\ \mathrm{d}_{t} a_{k} \end{pmatrix}.
  \endgroup
  \label{eq:astatecontrols}
\end{equation}
Here, the superscript on the quantum states $i \in \{0, 1\}$ acts as a label.
In standard QOC frameworks, the derivatives of the control fields
are obtained with finite difference methods, e.g.,
$\mathrm{d}_{t} a_{k} \approx (a_{k + 1} - a_{k}) / \Delta t$ \cite{leung2017speedup}.
Because ALTRO requires that cost functions do not use information from multiple time steps,
we make $\mathrm{d}^{2}_{t} a_{k}$ a decision variable and
numerically integrate coupled ODEs to obtain $\mathrm{d}_{t} a_{k}$, $a_{k}$, and $\int_{t} a_{k}$
so that we may penalize them in cost functions.
Similarly, the quantum states are obtained by numerically integrating
the TDSE \eqref{eq:tdse} with the fluxonium Hamiltonian \eqref{eq:hamiltonian}
and the given flux $a_{k}$. These numerical integration rules are implemented
in the discrete dynamics function for the problem, and they give rise to the
dynamics constraint \eqref{eq:dyn_con}.

Next, we outline the constraints for this problem.
Casting this problem in terms of a multi-state transfer problem, we fix as the initial states
 $\ket*{\psi^{0}_{1}} = \ket*{0}$, $\ket*{\psi^{1}_{1}} = \ket*{1}$
\eqref{eq:istate_con}.
The states at the final time step are constrained to be
the target states $\ket*{\psi^{i}_{N}} = \ket*{\psi^{i}_{T}} \equiv
U \ket*{\psi^{i}_{1}} \ \forall \ i$
\eqref{eq:tstate_con} where $U = X/2, Y/2, Z/2$ denotes the target gate.
Furthermore, we impose the normalization constraint
${\lvert \braket*{\psi^{i}_{k}}{\psi^{i}_{k}} \rvert}^{2} = 1 \ \forall \ i,k$
\eqref{eq:statenorm_con}
to ensure the solver does not take advantage of discretization errors in numerical integration.
For the flux,
we have the initial condition $\int_{t} a_{1} = \mathrm{d}_{t} a_{1} = 0$
\eqref{eq:ic_con}.
We also enforce the boundary condition $a_{1} = a_{N} = 0$ \eqref{eq:ic_con}, \eqref{eq:znf_con}
so the gates may be concatenated arbitrarily. 
We impose the zero net-flux constraint $\int_{t} a_{N} = 0$
\eqref{eq:znf_con}
which mitigates the inductive drift ubiquitous in flux-bias lines
\cite{battistel2019fast, krantz2019quantum, zhang2020universal}.
Additionally, the flux is constrained by $\lvert a_{k} \rvert \leq 0.5 \ \textrm{GHz}
\ \forall \ k$ \eqref{eq:amp_con} to ensure the two-level
approximation remains valid \eqref{eq:hamiltonian}. Above $0.5$ GHz,
the relationship between the energy levels and the flux becomes strongly non-linear.
All gates presented in this work satisfy these constraints to
a maximum violation of $\sim 10^{-8}$.

The cost function at each time step is
$\ell_{k}(\mathbf{x}_{k}, \mathbf{u}_{k}) = (\mathbf{x}_{k}
- \mathbf{x}_{T})^{T} Q_{k} (\mathbf{x}_{k} - \mathbf{x}_{T})
+ \mathbf{u}^{T}_{k} R_{k} \mathbf{u}_{k}$
where $Q_{k}$ and $R_{k}$ are diagonal matrices
of hyperparameters that assign weights to cost function contributions.
The $Q_{k}$ term
penalizes deviations from the target augmented state
$\mathbf{x}_{T} = (\psi^{0}_{T}, \psi^{1}_{T}, 0, 0, 0)^{T}$,
which is consistent with the constraints we have
imposed on $\ket*{\psi^{i}_{N}}$, $\int_{t} a_{N}$, and $a_{N}$.
Accordingly, this term penalizes the squared
difference of $\psi^{i}_{k}$ and $\psi^{i}_{T}$
and penalizes the norm of $\int_{t} a_{k}$, $a_{k}$, and $\mathrm{d}_{t} a_{k}$.
We penalize the squared difference of the final and target
quantum states, rather than their infidelities,
because the Hessian of the squared-difference cost function is diagonal -- which
makes matrix multiplications fast -- and we wish to optimize $Z/2$ gates,
which requires a metric that is sensitive to global phases for the initial
states $\ket*{0}$ and $\ket*{1}$.
Additionally, the $R_{k}$ term penalizes the norm of $\mathrm{d}^{2}_{t} a_{k}$.
Penalizing the norm of $\mathrm{d}^{2}_{t} a_{k}$ and $\mathrm{d}_{t} a_{k}$
makes $a_{k}$ smooth, which mitigates high-frequency AWG transitions.
Stated succinctly, the optimization problem takes the form:
\begin{mini!}[2] 
  {\substack{\mathbf{x}_1,\ldots,\mathbf{x}_N \\ \mathbf{u}_1,\ldots,\mathbf{u}_{N-1}}}
  {\sum_{k=1}^N {(\mathbf{x}_k-\mathbf{x}_{T})}^{T} Q_k (\mathbf{x}_k-\mathbf{x}_{T})
    + \sum_{k=1}^{N-1} {\mathbf{u}_k}^{T} R_k \mathbf{u}_{k}}{}{} \label{eq:costfun}
  \addConstraint{\mathbf{x}_{k+1}}{= \mathbf{f}(\mathbf{x}_k, \mathbf{u}_k) \ \forall \ k}
  \label{eq:dyn_con}
  \addConstraint{\ket*{\psi^{0}_{1}} = \ket*{0}, \ket*{\psi^{1}_{1}} = \ket*{1}}
  \label{eq:istate_con}
  \addConstraint{{\textstyle \int_{t}} a_{1} = a_{1} = \mathrm{d}_{t} a_{1} = 0}
  \label{eq:ic_con}
  \addConstraint{\ket*{\psi^{i}_{N}} = \ket*{\psi^{i}_{T}}
    \ \forall \  i} \label{eq:tstate_con}
  \addConstraint{{\textstyle \int_{t}} a_{N} = a_{N} = 0} \label{eq:znf_con}
  \addConstraint{{\lvert \braket{\psi^{i}_{k}}{\psi^{i}_{k}}
      \rvert}^{2} = 1 \ \forall \ i, k} \label{eq:statenorm_con}
  \addConstraint{|a_{k}| \leq 0.5 \ \textrm{GHz} \ \forall \ k.} \label{eq:amp_con}
\end{mini!}

Next, we remark on our problem formulation.
We put a cost function at all time steps
because it benefits the iLQR solving stage \cite{Jackson2020altroc};
although this may incentivize early achievement of the desired gate,
as in Ref. \cite{leung2017speedup}, we are primarily concerned
with achieving the gate at the final time step, which the target-state
constraint \eqref{eq:tstate_con} ensures.
Additionally, the target-state constraint
requires the final state to match the target state, including its global phase,
up to our chosen maximum constraint violation $\sim 10^{-8}$.
If we did not impose this constraint, the optimizer would be
allowed to sacrifice the closed-system gate error to achieve better
performance on the other cost functions, which is undesirable.
To enforce a constraint in standard QOC frameworks,
the prefactor for the constraint function is manually increased
between separate optimization instances until the constraint is satisfied
\cite{heeres2017implementing, leung2017speedup, reinhold2019controlling},
which becomes infeasible for more than one constraint.
ALM automates these prefactor updates to find
a solution that satisfies all of the given constraints.
Hence, ALTRO's ability to handle multiple constraints makes it
an attractive solver for QOC problems.

In extraordinarily difficult cases of
QOC, it may be impossible
to obey the physics of the system and achieve the desired gate \cite{abdelhafez2020universal},
i.e., the dynamics constraint \eqref{eq:dyn_con}
and the target-state constraint \eqref{eq:tstate_con} may be mutually unsatisfiable.
In this case, the prefactors for the constraint function terms
in the ALM objective will tend to infinity -- leading to numerical instability -- and the
optimization will not converge. To maintain a constrained approach in this situation,
the maximum constraint violation for the target-state constraint can be raised
to a level commensurate with the minimum acceptable gate error.

Finally, for ALTRO's first indirect stage,
the augmented states are obtained explicitly with the discrete
dynamics function, so the dynamics constraint and initial conditions
\eqref{eq:dyn_con}-\eqref{eq:ic_con} are satisfied by construction.
In this stage, the rest of the constraint functions \eqref{eq:tstate_con}-\eqref{eq:amp_con}
are added to the objective in their isomorphism-equivalent form \eqref{eq:isomorphism}.
Conversely, for the second direct stage, all of the constraints
\eqref{eq:dyn_con}-\eqref{eq:amp_con} are used to define
the projection onto the constraint manifold, and the objective is unmodified.
Hence, the quantum states become
free parameters that are adjusted to satisfy the TDSE.
Although the final solution's deviation from the TDSE is never
more than the maximum constraint violation,
we explicitly integrate the TDSE when reporting gate errors to ensure accuracy.
Exploring the benefit of direct optimization approaches
for QOC is an interesting direction for future work.

\begin{figure*}[ht]
  \begin{subfigure}{.315\textwidth}
    \includegraphics[width=\linewidth]{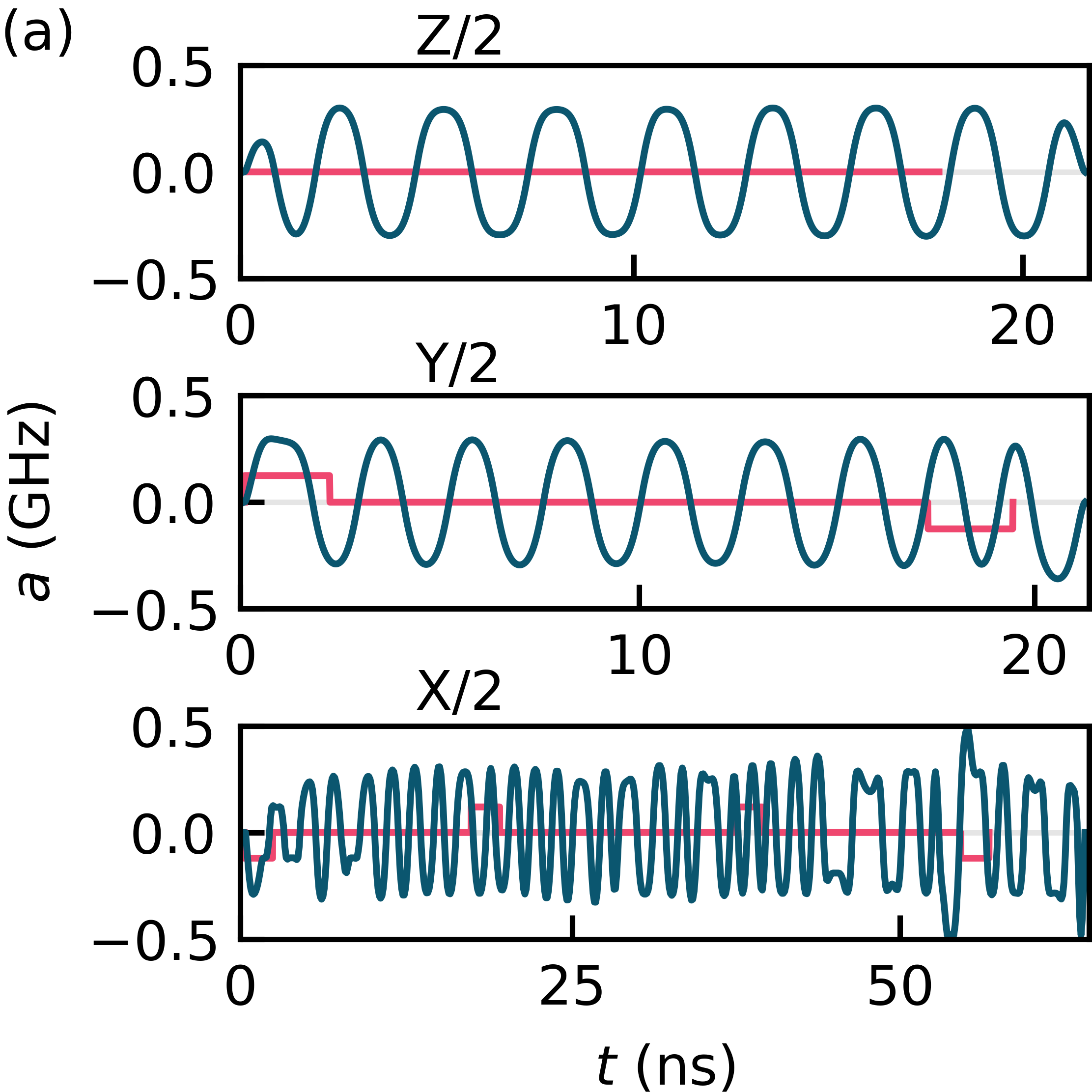}
    \caption{\label{fig:longitudea}}
  \end{subfigure}\hfill
  \begin{subfigure}{.23\textwidth}
    \includegraphics[width=\linewidth]{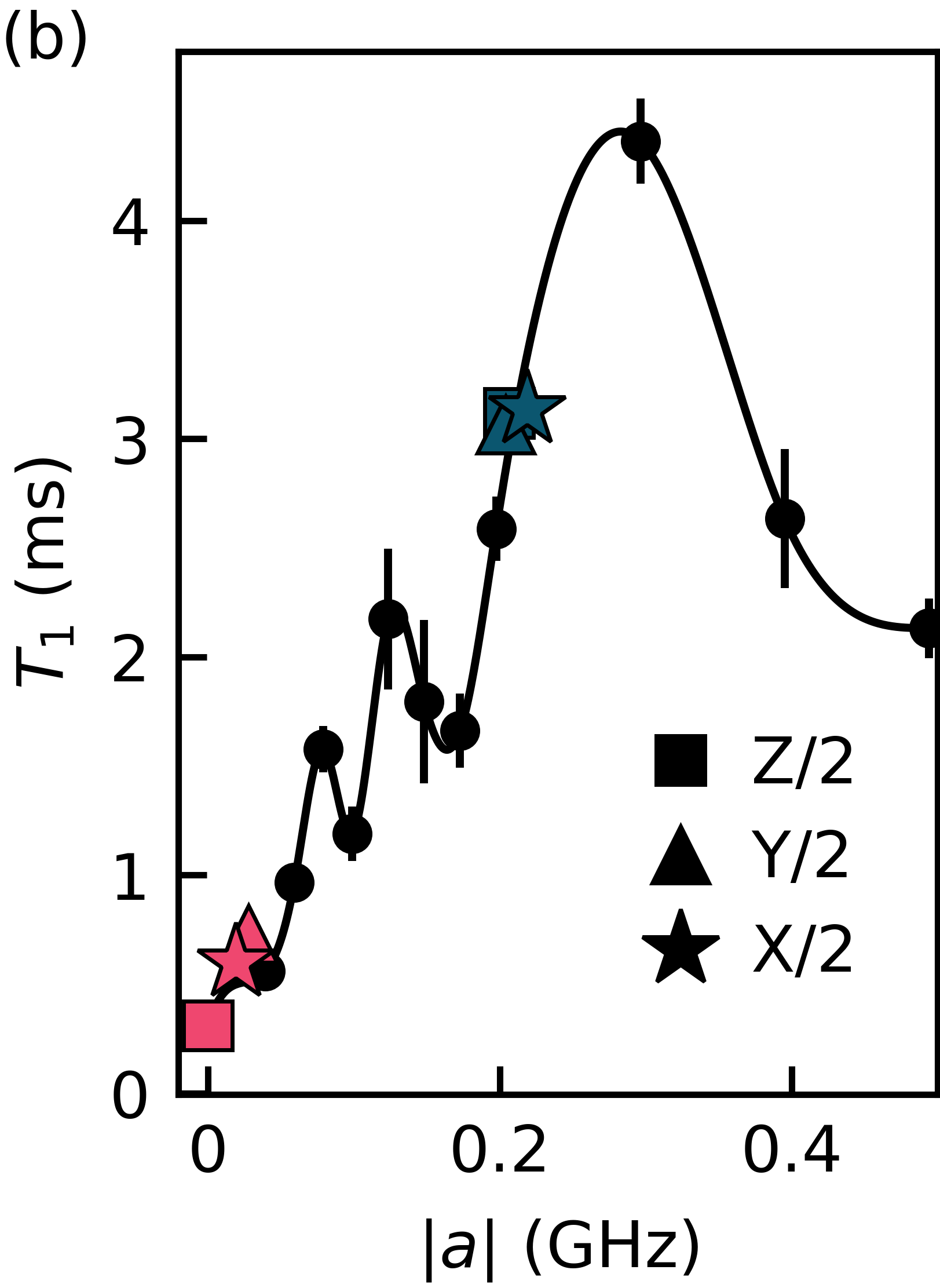}
    \caption{\label{fig:longitudeb}}
  \end{subfigure}\hfill
  \begin{subfigure}{.4\textwidth}
    \includegraphics[width=\linewidth]{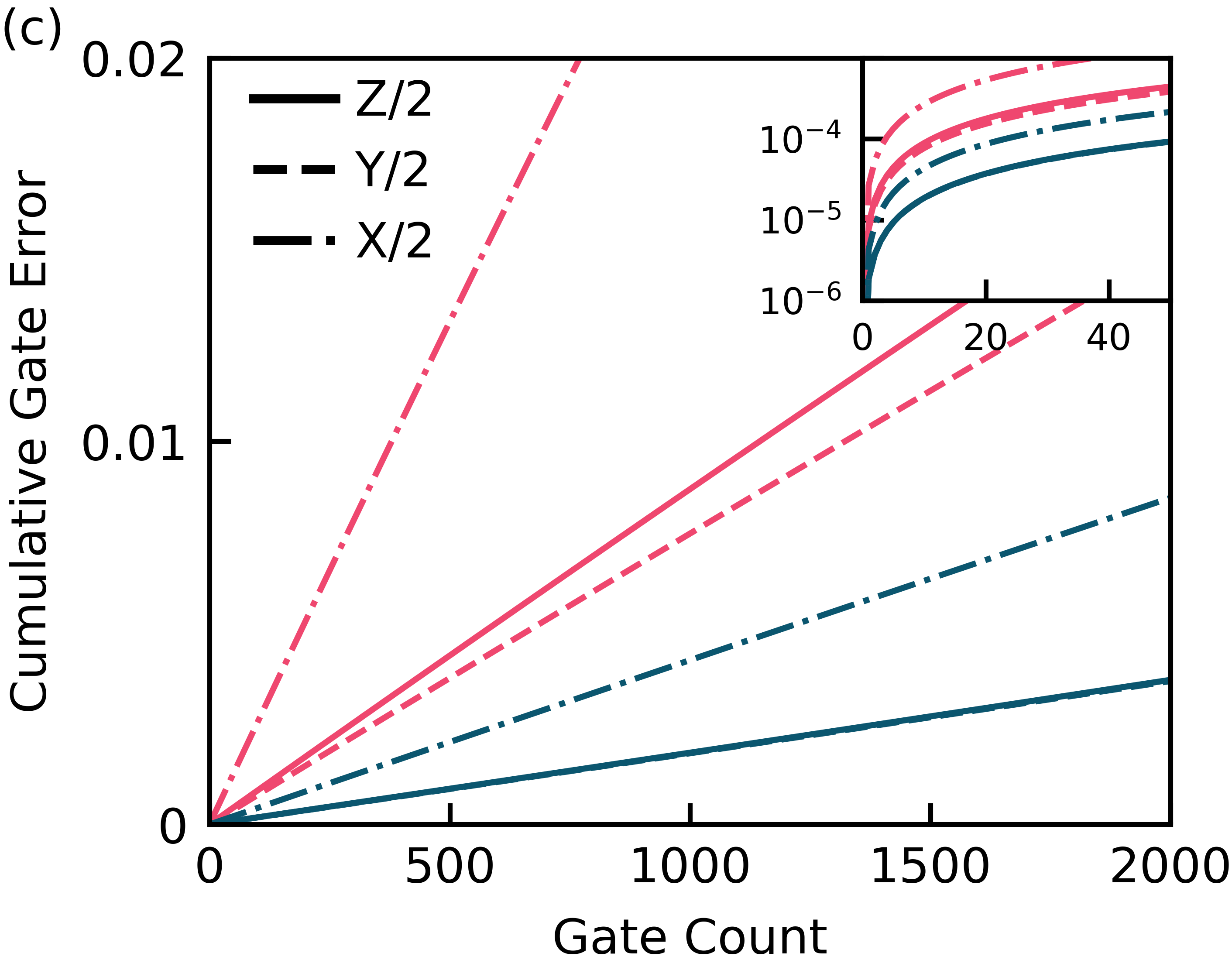}
    \caption{\label{fig:longitudec}}
  \end{subfigure}
  \caption{
    (a) Flux pulses for the numerical gates (dark blue)
    and the analytic gates (light pink).
    (b) $T_{1}$ interpolation function used in optimization. Circle markers
    indicate measured $T_{1}$ times. Non-circle markers
    are plotted at the time-averaged 
    absolute flux and the time-averaged $T_{1}$ time for each pulse.
    (c) Cumulative gate errors due to depolarization as a function of the
    number of gates applied.
    Cumulative gate errors for the numerical $Z/2$ and $Y/2$ gates
    are indistinguishable. Inset shows log-scaled cumulative gate errors
    for small gate counts.
  }
  \label{fig:longitude}
\end{figure*}

\section{Depolarization Mitigation\label{sec:longitude}}
In this section, we outline a method
for optimizing the flux to mitigate depolarization.
For many superconducting circuits, the depolarization time
$T_{1}$ is independent of the control parameters,
so the fastest possible gate incurs the least depolarization error
\cite{schulteherbruggen2011optimal}.
For the fluxonium, however, $T_{1}$ is strongly dependent on the flux.
We enable the optimizer to trade longer gate times
for longer $T_{1}$ times, or shorter $T_{1}$ times for shorter gate times,
by making the gate time a decision variable.
Additionally, previous work has modeled the gate error due to depolarization
by evolving density matrices under a master
equation \cite{rembold2020introduction, schulteherbruggen2011optimal},
or evolving a large number of states in a quantum trajectory approach
\cite{abdelhafez2019gradient}.
We avoid the increase in computational complexity required for these
techniques by penalizing the integrated depolarization rate in optimization.

The integrated depolarization rate is given by,
\begin{equation}
  D_{1}(t) = \int_{0}^{t} T_{1}^{-1}[a(t^{\prime})] dt^{\prime}.
\end{equation}
For the gates we consider here, where the gate time is small compared to $T_{1}$,
the integrated depolarization rate is proportional to the probability of a depolarization event.
Additionally, the integrated depolarization rate is a reasonable proxy for the gate error incurred
because depolarization errors are incoherent -- they increase
monotonically in time without interference.
The integrated depolarization rate is appended to the augmented state \eqref{eq:astatecontrols}
and its norm is penalized in the $Q_{k}$ term of the objective by setting
the corresponding element of the target augmented state to zero, see \eqref{eq:costfun}.
$T_{1}$ as a function of the flux is obtained by evaluating
a spline fit to experimental data, see Fig. \ref{fig:longitudeb}.

Alternatively, modeling the depolarization with a master equation approach would require adding
density matrices of size $n \times n$
to the augmented state, and a quantum trajectory approach
would require adding many states of size $n$ to the augmented state,
where $n$ is the dimension of the Hilbert space.
By contrast, the integrated depolarization rate is a single
real number; thus, the computational complexity of evaluating this
depolarization model does not scale
with the dimension of the Hilbert space.

To perform time-optimal control, we make
the duration between time steps
a decision variable \cite{howell2019altro}. 
The square root of the duration $\sqrt{\Delta t_{k}}$
is appended to the augmented control \eqref{eq:astatecontrols}
and its square $\lvert \Delta t_{k} \rvert$ is used
for integration in the discrete dynamics function.
Although we constrain the bounds of the duration between reasonable positive values to maintain
numerical stability, the optimizer may assign negative values to the
duration for intermediate optimization iterations, so this squaring
approach maintains positivity.

We analyze the effect of depolarization on
the $X/2$, $Y/2$, and $Z/2$ gates obtained with
our numerical method and the corresponding analytic gates presented in \cite{zhang2020universal}.
We use the Lindblad master equation to simulate $T_{1}$ dissipation for successive
gate applications, and compute the cumulative gate error
after each application, see Appendix \ref{appendix:longitude}.
The gate error reported in this text is the infidelity
of the evolved state and the target state averaged over 1000 pseudo-randomly
generated initial states.

The flux pulses for the numerical gates
are approximately periodic
with amplitudes $\sim 0.2 \textrm{GHz}$, see Fig. \ref{fig:longitudea}.
They are reminiscent of the analytically determined Floquet operations
for a fluxonium described in \cite{huang2021engineering}
and realized in \cite{mundada2020floquet}.
The numerical gate times are greater
than the analytic gate times, but the
numerical flux pulses
spend more time at larger flux values, achieving higher $T_{1}$ times on average,
see Fig. \ref{fig:longitudeb}.
The single-gate errors for both the analytic and numerical gates are
less than $10^{-4}$, which makes them sufficient for quantum error correction -- a
prerequisite for fault-tolerant quantum
computing \cite{aharonov2008fault, knill2005quantum, gottesman1997stabilizer}.
However, the numerical gates achieve single-gate errors
$\sim 5$ times less than those for the analytic gates,
which tracks closely with their relative improvement
on the integrated depolarization rate metric, see Appendix \ref{appendix:longitude}.
This advantage in single-gate errors corresponds to a
significant reduction in error correction resources
\cite{paetznick2014resource, suchara2013comparing}.
Furthermore, for successive gate applications,
the gate error due to depolarization is approximately linear
in the gate count, which we expect for $t \ll T_{1}$, see Fig. \ref{fig:longitudec}.
The gate error reduction for large gate counts is
important for noisy, intermediate-scale quantum (NISQ)
applications. These improvements are significant given the constraints
we have imposed on the gates,
and do not represent a fundamental limit to the optimization
methods we have employed.

\section{Robustness to Static Parameter Uncertainty \label{sec:static}}
We have formulated the QOC
problem as an \emph{open-loop} optimization problem, i.e.,
we do not incorporate feedback from the experiment into the optimization.
However, the precise device parameters will deviate from the parameters we use in optimization,
leading to poor experimental performance. We combat errors
of this form using robust control techniques,
making the state evolution insensitive
to parameter uncertainty. As an example,
we mitigate errors arising from the drift and finite measurement
precision of the qubit frequency, which modifies the fluxonium Hamiltonian
\eqref{eq:hamiltonian} by $f_{q} \rightarrow f_{q} + \delta f_{q}$.
We consider three robust control techniques to accomplish this task:
a sampling method, an unscented sampling method,
and a derivative method.

\subsection{Sampling Method}
The sampling method incentivizes the optimizer
to ensure that multiple copies of a state, each evolving
with a distinct value of the uncertain parameter, achieve
the same target state. Variants of this technique have been proposed
in the context of QOC
\cite{allen2019robust, khaneja2005optimal,
  reinhold2019controlling, rembold2020introduction}.
For each initial state,
we add two sample states $\ket*{\psi^{\pm}}$
to the augmented state \eqref{eq:astatecontrols}. The discrete dynamics
function is modified
so the sample states evolve under the fluxonium Hamiltonian \eqref{eq:hamiltonian}
with $f_{q} \rightarrow f_{q} \pm \sigma_{f_{q}}$ for a fixed
 standard deviation $\sigma_{f_{q}}$ of the qubit frequency, acting as a hyperparameter.
We penalize the infidelities of the sample states with respect to the target state
by adding a cost function to the objective of the form
$\sum_{k, \pm} b_{k} (1 - {\lvert \braket*{\psi_{T}}{\psi^{\pm}_{k}} \rvert}^{2})$
where $b_{k}$ is a constant we supply.
For this method, the standard orthonormal basis states are an insufficient choice
for the initial states. As an example, a $Z/2$ gate achieved by idling
at the flux frustration point ($a_{k} = 0 \ \forall \ k$)
will be robust to qubit frequency detunings for the initial states $\ket{0}$
or $\ket{1}$ because the infidelity metric is insensitive to global phases,
but this gate will not be robust for any other initial states.
Therefore, we choose the four initial states $\{\ket{0}, \ket{1}, (\ket{0} + i\ket{1}) / \sqrt{2},
(\ket{0} - \ket{1}) / \sqrt{2}\}$ \cite{chow2009randomized},
whose outer products span the operators on the Hilbert space,
and  we refer to them as the operator basis.

\subsection{Unscented Sampling Method}
Whereas the sampling method penalizes the deviations of the sample states
from the target state, the unscented sampling method
penalizes the deviations of the sample states from the nominal state
\cite{howell2020direct, lee2013sigma,
  thangavel2020robust}. Accordingly, the cost function we add
to the objective takes the form
$\sum_{k, j} c_{k} (\psi^{j}_{k} - \psi_{k})^{T}
(\psi^{j}_{k} - \psi_{k})$, where $c_{k}$ is a
constant we supply, $\psi_{k}$ is
the evolved initial state (nominal state), and $\psi^{j}_{k}$ is a sample state
that evolves under a modified Hamiltonian similar to that in the sampling method.
The sample states are chosen to encode a unimodal distribution over
the $2n$ elements of the nominal state, modeling the uncertainty in the state
as a result of the uncertainty in the parameter. We use the unscented transform
\cite{julier2004unscented, uhlmann1995dynamic}
to accurately propagate the mean and covariance of this distribution between
time steps, or equivalently, through the transformation of the TDSE \eqref{eq:tdse}.
Unlike the sampling method, the cost function for the unscented sampling method
is sensitive to global phases. Accordingly, we
do not observe a performance increase when
using more than one initial state. 
A detailed procedure for the unscented transformation is given
in Appendix \ref{appendix:unscented}.

\subsection{Derivative Method}
The derivative method penalizes the sensitivity of the state
to the uncertain parameter, which is encoded in the $l$\textsuperscript{th}-order
state derivative $\ket*{\partial_{f_{q}}^{l} \psi} \equiv \partial_{f_{q}}^{l} \ket*{\psi}$.
In the $m$\textsuperscript{th}-order
derivative method, we append all state derivatives of order $1, \dots, m$
to the augmented state \eqref{eq:astatecontrols}
for each initial state.
We obtain the state derivatives at each time step by performing forward-mode
differentiation on the TDSE \eqref{eq:tdse}.
For example, the dynamics for the $1$\textsuperscript{st}-order derivative method are:
\begin{align}
  i \hbar \frac{d}{dt} \ket{\psi} &= H \ket{\psi},\\
  i \hbar \frac{d}{dt} \ket{\partial_{f_{q}}\psi} &=
  H \ket{\partial_{f_{q}} \psi} +
  (\partial_{f_{q}} H) \ket{\psi}.
  \label{eq:d1dyn}
\end{align}
We integrate the coupled ODEs with exponential
integrators, see Appendix \ref{appendix:derivative}.
While the state $\ket*{\psi}$ has unit norm,
the state derivatives $\ket*{\partial^{l}_{f_{q}} \psi}$ need not, as is evident
from the non-unitary dynamics \eqref{eq:d1dyn}.
We penalize the norms of the isomorphism-equivalent state derivatives
in the $Q_{k}$ term of the objective by setting the corresponding elements
of the target augmented state to zero, see \eqref{eq:costfun}. Intuitively, this corresponds to penalizing
the sensitivity of each state element to the uncertain parameter. As was the case for
the unscented sampling
method, we do not observe a performance increase when using more than one initial state
for the derivative method.
We present the runtimes of our implementations of the three robust control methods
in Appendix \ref{appendix:time}.

\begin{figure*}[ht]
  \begin{subfigure}{.315\textwidth}
    \includegraphics[width=\linewidth]{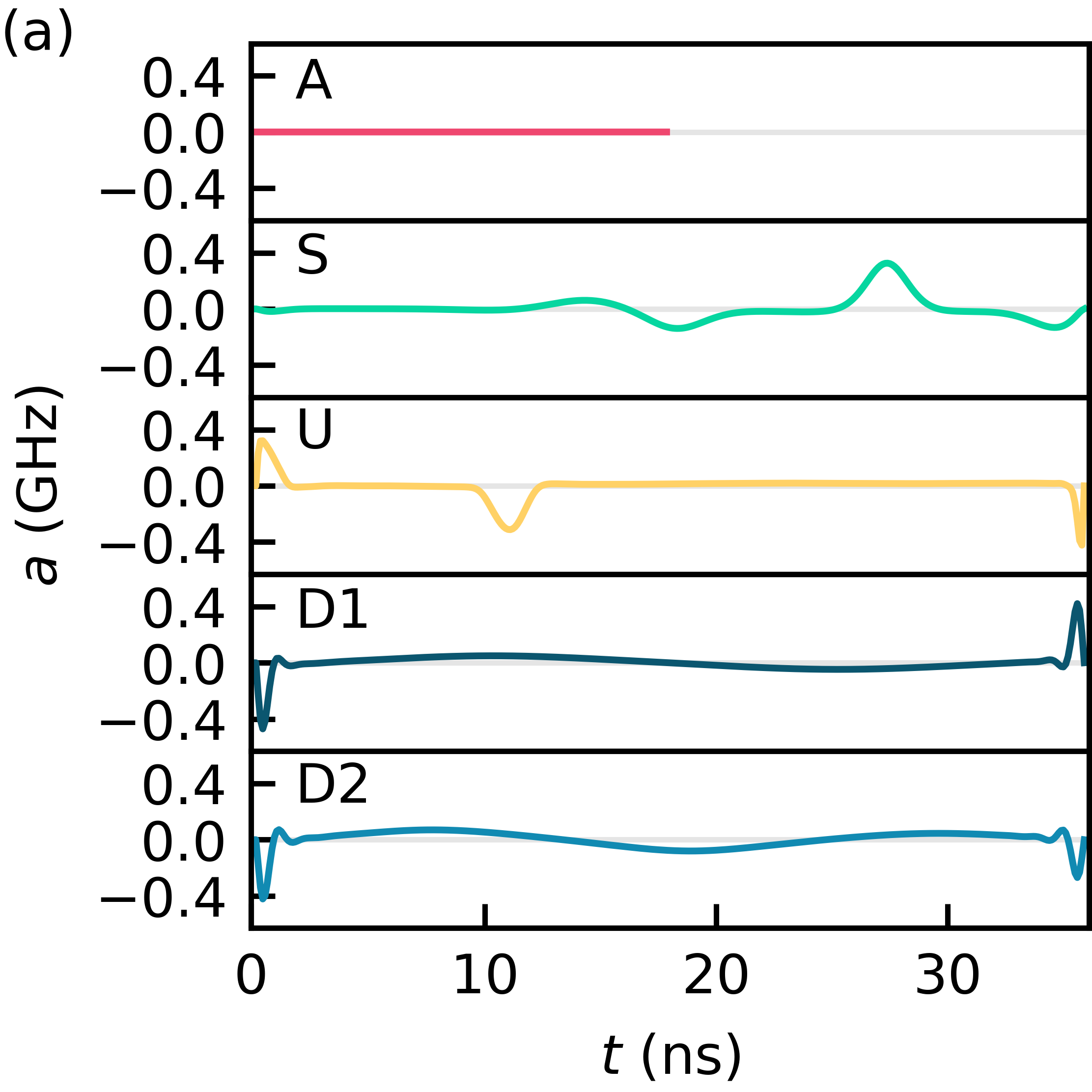}
    \caption{\label{fig:statica}}
  \end{subfigure}\hfill
  \begin{subfigure}{.4\textwidth}
    \includegraphics[width=\linewidth]{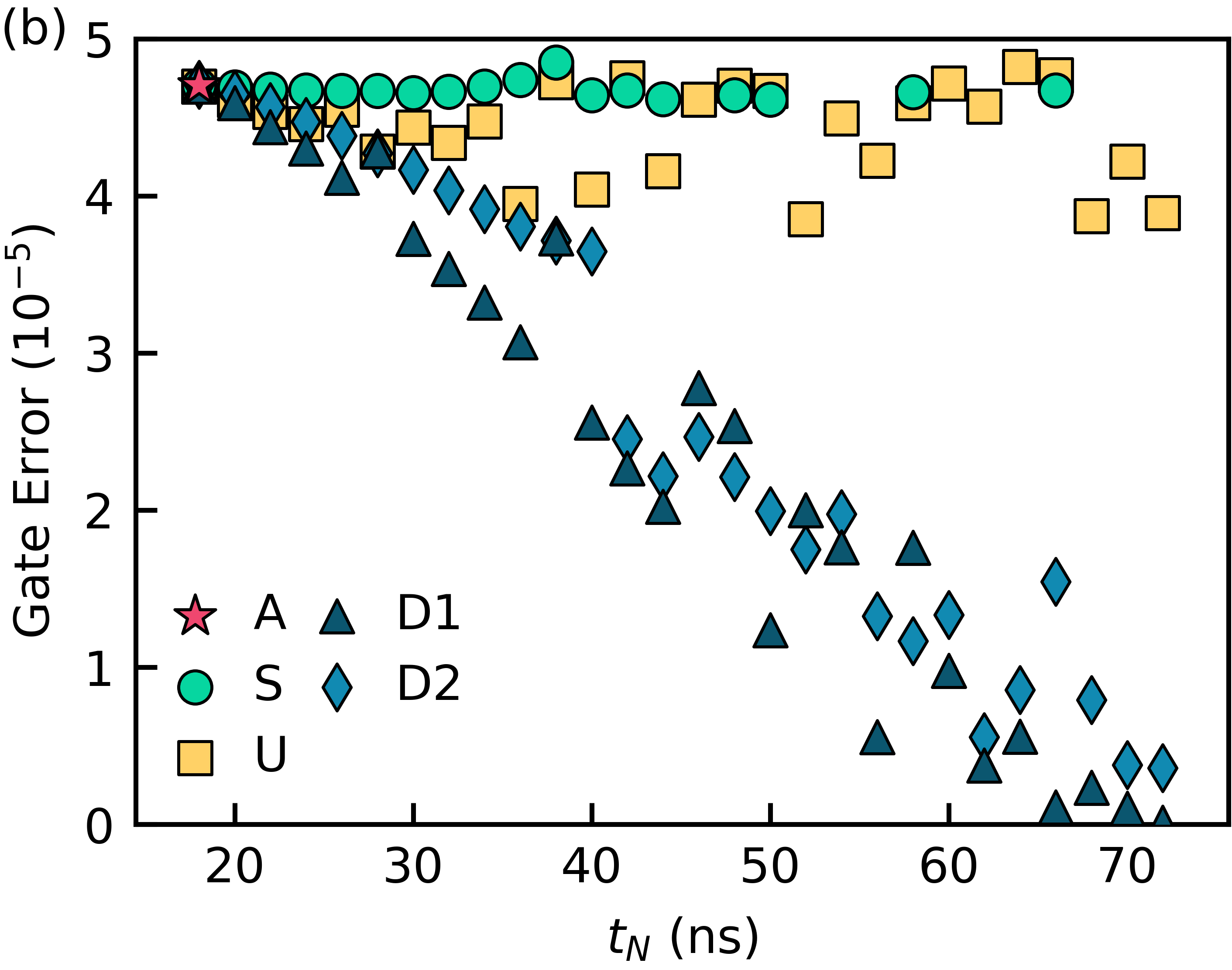}
    \caption{\label{fig:staticb}}
  \end{subfigure}\hfill
  \begin{subfigure}{.23\textwidth}
    \includegraphics[width=\linewidth]{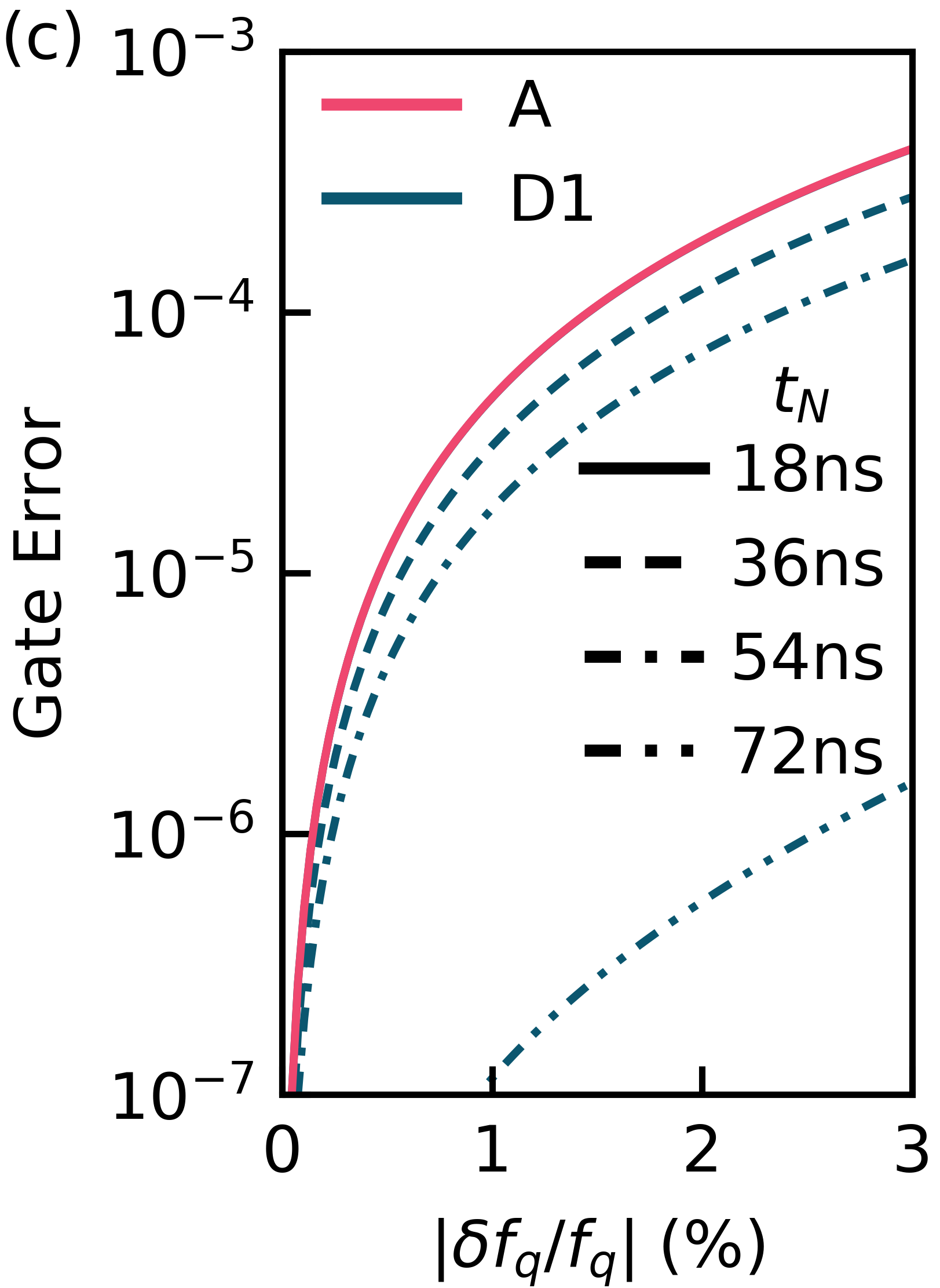}
    \caption{\label{fig:staticc}}
  \end{subfigure}
  \caption{
    (a) Flux pulses for $Z/2$ gates robust to qubit frequency detunings constructed with the
    analytic (A), sampling (S), unscented sampling (U), and the 1\textsuperscript{st}-
    and 2\textsuperscript{nd}-order derivative methods (D1, D2). The flux pulses shown
    for the sampling, unscented sampling, and derivative methods are optimized
    for twice the gate time of the analytic gate.
    (b) Single-gate error at a one-percent qubit frequency detuning as
    a function of the gate time. Missing
    data points represent gates with a gate error greater than $5 \cdot 10^{-5}$.
    (c) Single-gate error as a function of the qubit frequency detuning.
    The gate errors for the analytic and 1\textsuperscript{st}-order derivative
    methods are shown for gate times which are multiples of $1 / 4 f_{q} \sim 18 \textrm{ns}$.
    The gate errors for the two methods are
    indistinguishable at the gate time $18 \textrm{ns}$.
  }
  \label{fig:static}
\end{figure*}

\subsection{Comparison}
We examine the gate errors due to a static qubit frequency
detuning for the $Z/2$ gates obtained with the robust control techniques
and the analytic $Z/2$ gate.
To compute the gate error,
an initial state is evolved
under the fluxonium Hamiltonian \eqref{eq:hamiltonian}
two separate times with the transformations
$f_{q} \rightarrow f_{q} \pm \delta f_{q}$
at the stated qubit frequency detuning $\delta f_{q}$.
The reported gate error is the infidelity of
the evolved state and the target state averaged over
the two transformations for each of $1000$ pseudorandomly
generated initial states.
We set $\sigma_{f_{q}}/f_{q} = 1\%$
for the sampling and unscented sampling
methods.

The analytic gate corresponds to
idling at the flux frustration point $a_{k} = 0 \ \forall \ k$, see Fig.
\ref{fig:statica}. Its gate time $1 / 4 f_{q} \sim 18\textrm{ns}$
is the shortest possible for a $Z/2$ gate on the device.
The gate's erroneous rotation angle
$2 \pi \delta f_{q} / 4 f_{q}$ is linear in the
qubit frequency detuning, resulting in a gate error that is quadratic
in the detuning.
At a one-percent detuning $\lvert \delta f_{q} / f_{q} \rvert = 1\%$,
the gate error is $\sim 4.7 \cdot 10^{-5}$,
which is sufficient for quantum error correction.

For the sampling method, the gate error at a one-percent qubit frequency detuning
does not decrease substantially over the
range of gate times, and begins to increase above $5 \cdot 10^{-5}$ for gate times
greater than $\sim 50$ns, see Fig. \ref{fig:staticb}.
Optimization results for the sampling method reveal that it is typically
able to achieve a high fidelity for
one sample $\ket{\psi^{\pm}}$,
but not the other $\ket{\psi^{\mp}}$, indicating that it is difficult for the optimizer
to make progress on both objectives.
For the unscented sampling method,
the gate error at a one-percent detuning
does not decrease substantially 
over the gate times, but it does reach
a minimum of $\sim 3.9 \cdot 10^{-5}$
near fractions of the Larmor period: $2/4f_{q} \sim 36\textrm{ns}$,
$3/4f_{q} \sim 54\textrm{ns}$, and $4/4f_{q} \sim 72\textrm{ns}$.

The two derivative methods converge on qualitatively similar flux pulses that
idle near the flux frustration point and use fast triangle movements at the boundaries,
similar to the flux pulse produced by the unscented sampling method.
For both derivative methods, the gate error at a one-percent qubit frequency detuning
decreases super-linearly in the gate time.
For the 1\textsuperscript{st}-order method, the gate error at a one-percent detuning 
reaches $10^{-7}$ at the Larmor period $1 / f_{q} \sim 72$ns,
see Fig. \ref{fig:staticc}.
This result mimics the
ability of composite pulses to mitigate parameter uncertainty errors to arbitrary
order with sufficiently many pulses \cite{merrill2014progress}.
It is difficult to choose an appropriate composite pulse
for the problem studied here due to our Hamiltonian and experimental constraints.
A comparison between composite pulses and numerical techniques could
be an interesting topic for future study.

Furthermore, the ability to perform
$Z$-type gates in any given time is critical
for synchronizing phases in multi-qubit experiments,
where the qubits have distinct
frequencies. Notably, the analytic gate studied here cannot be extended
to gate times other than $1/4f_{q}$. 
We can find gates using the numerical methods at
all gate times at and above $1/4f_{q}$, see Fig. \ref{fig:staticb}.
These numerical methods offer an effective scheme for synchronizing
multi-qubit experiments.

\begin{figure*}[ht]
  \begin{subfigure}{.4\textwidth}
    \includegraphics[width=\linewidth]{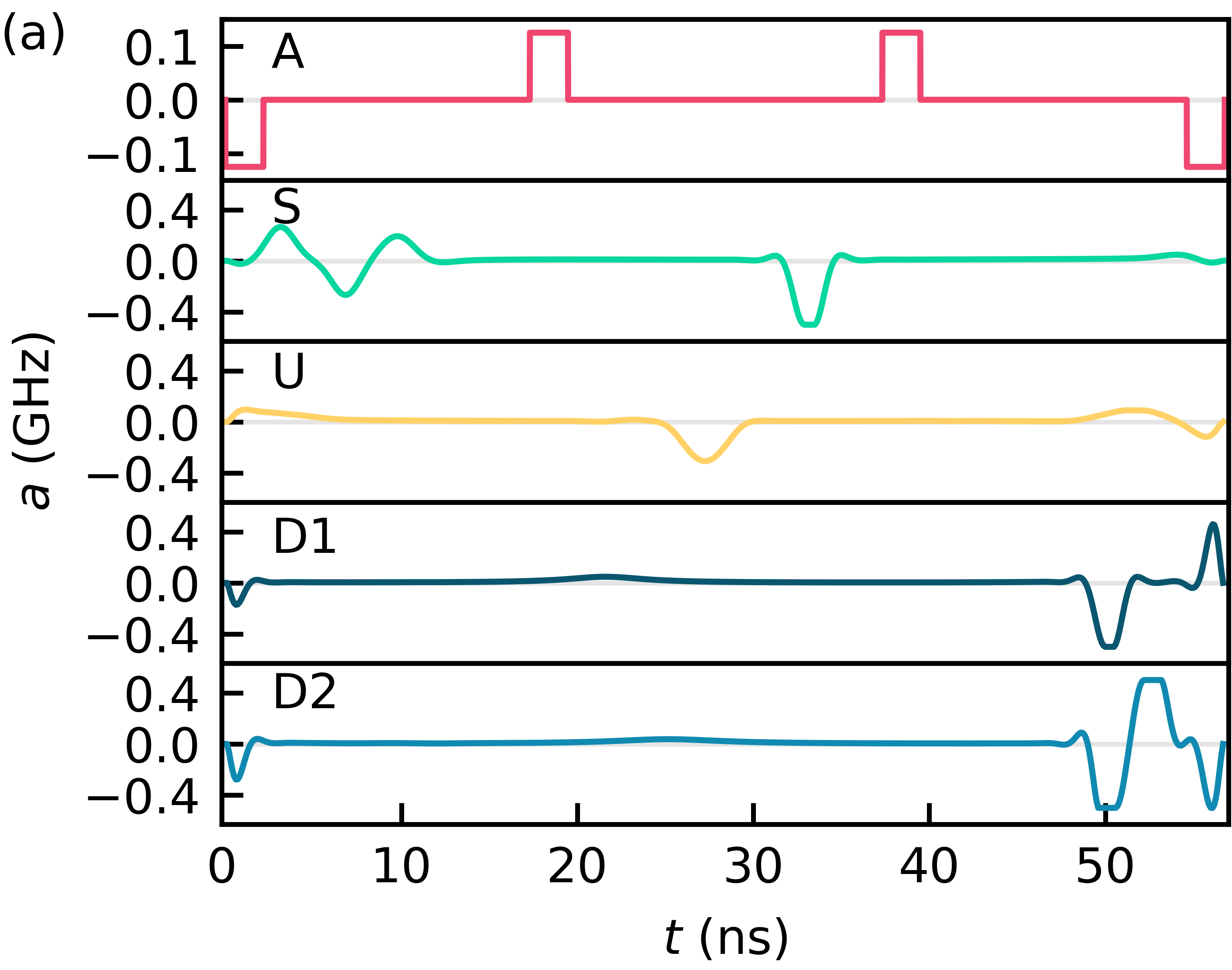}
    \caption{}
    \label{fig:stochastica}
  \end{subfigure}\hspace{0.05\textwidth}
  \begin{subfigure}{.4\textwidth}
    \includegraphics[width=\linewidth]{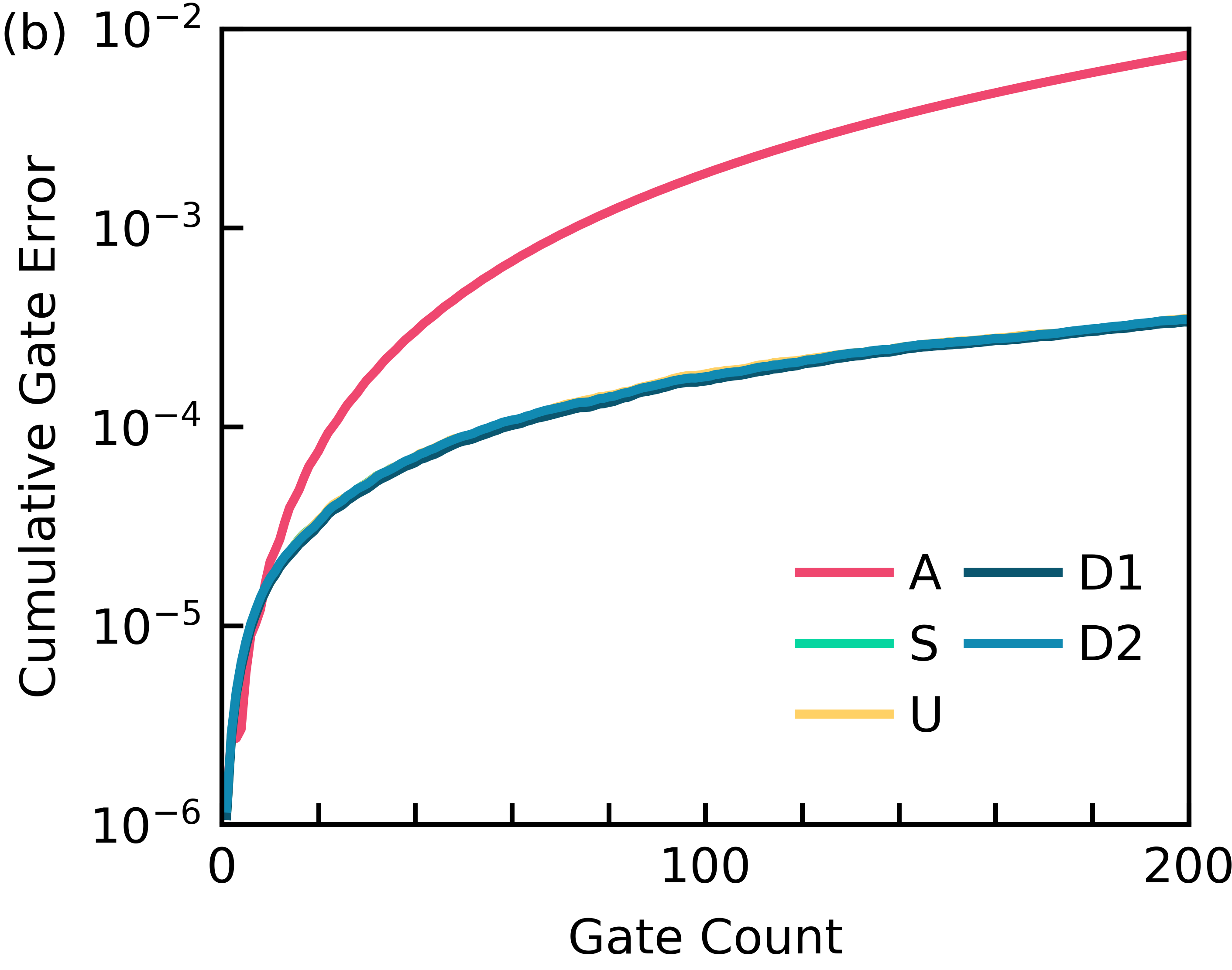}
        \caption{}
    \label{fig:stochasticb}
  \end{subfigure}
  \caption{
    (a) Flux pulses for $X/2$ gates robust to flux noise
    constructed with the analytic (A),
    sampling (S), unscented sampling (U), and the 1\textsuperscript{st}-
    and 2\textsuperscript{nd}-order derivative methods (D1, D2).
    (b) Cumulative gate error due to 1/$f$ flux noise for
    successive gate applications. The cumulative gate errors for the
    sampling, unscented sampling, and the derivative methods are indistinguishable.
  }
  \label{fig:stochastic}
\end{figure*}

\section{Robustness to Time-Dependent Parameter Uncertainty \label{sec:stochastic}}
An additional source of experimental error arises from time-dependent
parameter uncertainty. For many flux-biased and inductively-coupled
superconducting circuit elements, magnetic flux noise is the dominant
source of coherent errors \cite{bialczak20071f, kakuyanagi2007dephasing,
  kumar2016origin, yoshihara2006decoherence}. Flux noise
modifies the fluxonium Hamiltonian \eqref{eq:hamiltonian}
by $a(t) \rightarrow a(t) + \delta a(t)$ where $\delta a(t)$ is the flux noise.
The spectral density of flux noise is observed to
follow a 1/$f$ distribution
\cite{bialczak20071f, koch2007model, kakuyanagi2007dephasing,
  kumar2016origin, yoshihara2006decoherence, yoshihara2010correlated,
  zhang2020universal},
so the noise is dominated by low-frequency components.
The analytic gate considered here
takes advantage of the low-frequency characteristic and
treats the noise as quasi-static, performing a generalization of the spin-echo
technique to compensate for erroneous drift \cite{hahn1952spin, meiboom1958modified}.

We modify the robust control techniques presented
in the previous section to combat 1/$f$ flux noise.
The unscented sampling method is modified so that the sample states
are subject to 1/$f$ flux noise. The noise
is generated by filtering white noise sampled from a standard
normal distribution with a finite impulse response filter \cite{saspweb2011}.
The noise is then scaled by the 
flux noise amplitude of our device $A_{\Phi} = 5.21 \mu \Phi_{0} \implies
\sigma_{a} = 2.5 \cdot 10^{-5} \textrm{GHz}$.
In principle, we could modify the sampling method
similarly; however, we choose to subject the sample states
to static noise
$a(t) \rightarrow a(t) \pm \sigma_{a}$
for comparison. The derivative methods require no algorithmic modification
from the static case, but the TDSE is now differentiated with respect
to $a(t)$ instead of $f_{q}$ as in \eqref{eq:d1dyn}.

We analyze the gate errors due to 1/$f$ flux noise for
the $X/2$ gates constructed with the robust control techniques
and the analytic $X/2$ gate. To compute the gate error,
we evolve an initial state
under the fluxonium Hamiltonian \eqref{eq:hamiltonian}
where the optimized flux is modified $a(t) \rightarrow a(t) + \delta a(t)$.
We generate the flux noise as
we described for the unscented sampling method.
The reported gate error is the infidelity
averaged over $1000$ pseudorandomly generated initial states,
each of which is subject to a distinct pseudorandomly
generated flux noise instance.
To observe the effect of interfering coherent errors,
we simulate successive applications of the gate constructed by each method;
we compute the cumulative gate error
after each application, see Fig. \ref{fig:stochastic}.
Both the analytic
and numerical gates yield single-gate errors
sufficient for quantum error correction.
Despite converging on qualitatively different solutions, the
numerical gates perform similarly in the concatenated
gate application comparison. Their gate errors
after $200$ gate applications $\sim 11 \mu\textrm{s}$ are
two orders of magnitude less than the gate error produced by the analytic gate.
1/$f$ flux noise is a significant source of coherent errors in NISQ applications,
and these numerical techniques offer effective avenues to mitigate it.

\section{Conclusion}
We have introduced state-of-the-art trajectory optimization
techniques in the context of quantum optimal control, enabling
us to achieve tight tolerances for multiple constraints on the
control fields and quantum states. Using these capabilities,
we have mitigated decoherence and
achieved robustness to parameter uncertainty
errors on a superconducting fluxonium qubit.
We have proposed a scheme for suppressing
depolarization with time-optimal
control and the integrated depolarization rate model.
The computational complexity of evaluating this model is
independent of the dimension of the Hilbert space, enabling
inexpensive optimization on high-dimensional quantum systems.
We have also proposed the derivative method for robust control which achieves
superlinear gate error reductions in the gate time for the static parameter
uncertainty problem we studied.
We have shown that the derivative, sampling, and unscented sampling methods
can mitigate 1/$f$ flux noise errors -- which
dominate coherent errors for flux controlled qubits.
These robust control techniques can be applied
to any Hamiltonian,
allowing experimentalists in all domains to engineer robust
operations on their quantum systems.
Furthermore, they can be used to achieve the low gate errors
required for fault-tolerant quantum computing applications. Our
implementations of the techniques described in this work are available
at \url{https://github.com/SchusterLab/rbqoc}.

\begin{acknowledgments}
  We thank Helin Zhang for experimental assistance
  and Taylor Howell, Tanay Roy, Colm Ryan, and Daniel Weiss for useful discussions.
  This work was made possible by many open source software projects,
  including but not limited to:
  DifferentialEquations.jl \cite{rackauckas2017differentialequations},
  Distributions.jl \cite{besancon2019distributions},
  ForwardDiff.jl \cite{revelsLubinPapamarkou2016},
  Matplotlib \cite{hunter2007matplotlib},
  NumPy \cite{harris2020array},
  TrajectoryOptimization.jl \cite{howell2019altro},
  and Zygote.jl \cite{innes2018don}.
  This work is funded in part by EPiQC, an NSF Expedition in Computing, under grant CCF-1730449.
  This work was supported by the Army Research Office under Grant No. W911NF1910016.
\end{acknowledgments}

\appendix
\section{Depolarization \label{appendix:longitude}}
We comment on the depolarization metrics and then give
our procedure for integrating the Lindblad master equation.
The integrated depolarization rate and the gate error due to
depolarization
are compared in Table \ref{tab:longitude} for the
numerical experiment described in Sec. \ref{sec:longitude}.
The ratio of the value obtained on the metric with the analytic technique
to the value obtained with the numerical technique
is similar across the two metrics.

\begin{table}[ht]
  \begin{tabular}{| c | c | c | c | c | c | c | c | c |}
    \hline
    \multirow{2}{*}{Gate} &
    $D_{1\textrm{A}}$ &
    $D_{1\textrm{N}}$ &
    \multirow{2}{*}{$\dfrac{ D_{1\textrm{A}} }{ D_{1\textrm{N}} }$} &
    $\textrm{GE}_{\textrm{A}}$ &
    $\textrm{GE}_{\textrm{N}}$ &
    \multirow{2}{*}{$\dfrac{ \textrm{GE}_{\textrm{A}} }{ \textrm{GE}_{\textrm{N}} }$}\\
    &
    $(10^{-5})$ &
    $(10^{-5})$ &
    &
    $(10^{-5})$ &
    $(10^{-5})$ &\\
    \hline
    Z/2 & 5.745 & 1.149 & 5.000 & 0.888   & 0.185  & 4.791\\
    \hline
    Y/2 & 5.253 & 1.157 & 4.540 & 0.770 & 0.186   & 4.132\\
    \hline
    X/2 & 16.251 & 2.660 & 6.109 & 2.674 & 0.432  & 6.200\\
    \hline
  \end{tabular}
  \caption{
    Single-gate integrated depolarization rate ($D_{1}$)
    and single-gate error due to depolarization (GE).
    Values are reported for the analytic (A) and numerical (N) gates.
  }
  \label{tab:longitude}
\end{table}

We employ the Lindblad master equation
to compute the gate error due to depolarization.
This equation takes the form:
\begin{equation}
  \frac{d}{dt} \rho = -\frac{i}{\hbar} [H, \rho]
  + \sum_{i} \gamma_{i} (L_{i} \rho L_{i}^{\dagger}
  - \frac{1}{2} \{L_{i}^{\dagger} L_{i}, \rho\}),
\end{equation}
For depolarization, $\gamma_{\pm} = T_{\pm}^{-1}$,
$L_{\pm} = \sigma^{\pm} \equiv (\sigma_{x} \pm i \sigma_{y})/2$.
Our device operates in the regime where $hf \ll k_{\textrm{B}}T$ such that
$T_{+} = T_{-} = 2 T_{1}$, where $T_{1}$ is obtained at each time step
from the spline shown in Fig. \ref{fig:longitudeb}.
We obtain the $T_{1}$ values in this spline
by driving the qubit at the desired flux bias
and monitoring the resultant decay. For more details
on these measurements, consult Ref. \cite{zhang2020universal}.
Because $T_{1}$ depends on the flux, so do
the decay rates $\gamma_{\pm}$.
Integrating the master equation with time-dependent decay rates
provides a heuristic for how gates might  perform in
the experiment. 
This procedure may not be strictly correct
when decay rates change significantly on the time scale of the relaxation time,
which is the regime we are operating in. Standard
derivations of the Lindblad master equation do not account for
time-dependent decay rates \cite{manzano2020a}. A more thorough
treatment of this regime in future work would unlock new insights for
quantum computing platforms where decoherence is strongly
dependent on the control parameters.

In order to use exponential integrators, we employ
the vector (Choi-Jamiolkowski) isomorphism \cite{Landi2018},
\begin{equation}
  \frac{d}{dt} \textrm{vec}({\rho}) = \hat{\mathcal{L}}\, \textrm{vec}({\rho})
  \label{eq:veclindblad},
\end{equation}
\begin{equation}
  \begin{aligned}
    \hat{\mathcal{L}} &= -i(\openone \otimes H - H^{T} \otimes \openone)\\
    &+ \sum_{i} \gamma_{i}
    (L_{i}^{*} \otimes L_{i} - \frac{1}{2} (\openone \otimes L_{i}^{\dagger}L_{i}
    - L_{i}^{T}L_{i}^{*} \otimes \openone)),
  \end{aligned}
\end{equation}
where $\rho = \sum_{i, j} \alpha_{ij} \ket{i}\bra{j}$
and $\textrm{vec}(\rho) = \sum_{i, j} \alpha_{ij} \ket{i} \otimes \ket{j}$.
Because the flux is constant between time steps
due to our numerical discretization,
the Hamiltonian and decay rates are also constant
between time steps.
Therefore, the exact solution to \eqref{eq:veclindblad} is,
\begin{equation}
  \textrm{vec}(\rho_{k + 1}) = {\exp}{\textstyle(}\Delta t_{k}
  \hat{\mathcal{L}}_{k}{\textstyle)} \textrm{vec}(\rho_{k}).
\end{equation}
The vector isomorphism transforms $(n \times n) \times (n \times n)$
matrix-matrix multiplications to $(n^{2} \times n^{2}) \times n^{2}$ matrix-vector
multiplications. For small $n$, we find that it is
faster to use an exponential integrator on the vectorized equation than to perform
Runge-Kutta on the unvectorized equation.
The latter requires decreasing the interval $\Delta t_{k}$
to maintain accuracy, resulting in more time steps.

\section{Unscented Sampling Method}
\label{appendix:unscented}
In this section, we outline the full unscented sampling procedure.
We consider a state $\psi \in \mathbb{R}^{2n}$, 
an uncertain set of parameters $\lambda\in \mathbb{R}^{d}$, and discrete dynamics
$\psi_{k + 1} = f(\psi_{k}, \lambda_{k})$.
The nominal initial state is given by $\bar{\psi}_{1}$ with an associated
covariance matrix $P_{1} \in \mathbb{S}_{++}^{2n}$
which describes the uncertainty in the initial state.
We use the notation $\mathbb{S}_{++}^{m}$ to denote the
set of real, symmetric, and positive-definite $m \times m$ matrices.
By the positive-definite requirement, $P_{1}$ must be non-zero even if the state-preparation error
is negligible.
The uncertain parameter has zero-mean
and its distribution is given by the covariance matrix
$L_{k} \in \mathbb{S}_{++}^{d}$ at time step $k$. The zero-mean assumption
is convenient for deriving the update procedure. A non-zero mean can be encoded
in the discrete dynamics function $f(\psi_{k}, \lambda_{k})$.

The initial $4n + 2d$ sample states and initial $4n + 2d$
uncertain parameters are sampled from the initial distributions,
\begin{equation}\label{eq:uupdate}
  \begin{pmatrix} \psi_{1}^{j} \\ \lambda_{1}^{j} \end{pmatrix} =
  \begin{pmatrix} \bar{\psi}_{1} \\ 0\end{pmatrix}
    \pm \beta \sqrt{\begin{pmatrix} P_{1} & 0\\ 0 & L_{1}\end{pmatrix}}^{\; j}.
\end{equation}
Here, $\beta$ is a hyperparameter that controls the spacing of the covariance contour.
The $(\pm)$ is understood to take $(+)$ for $j \in \{1, \dots, 2n + d\}$ and $(-)$ for
$j \in \{2n + d + 1, \dots, 4n + 2d\}$. We use the Cholesky factorization
to compute the square root of the
joint covariance matrix, though other methods
such as the principal square root may be employed.
The superscript on the matrix square root indicates the $j$\textsuperscript{th}
column (mod $2n + d$) of the lower triangular Cholesky factor.
Then, the sample states are normalized,
\begin{equation}\label{eq:unormalize}
  \psi_{1}^{j} \rightarrow \frac{\psi_{1}^{j}}{\sqrt{{\psi_{1}^{j}}^{T} \psi_{1}^{j}}}.
\end{equation}
The sample states are propagated to the next time step,
\begin{equation}\label{eq:upropagate}
  \psi^{j}_{2} = f(\psi^{j}_{1}, \lambda^{j}_{1}).
\end{equation}
The mean and covariance of the sample states are computed,
\begin{align}
  \bar{\psi}_{2} &= \frac{1}{4n + 2d} \sum_{j = 1}^{4n + 2d} \psi_{2}^{j},\\
  P_{2} &= \frac{1}{2 \beta^{2}} \sum_{j = 1}^{4n + 2d}
  (\psi^{j}_{2} - \bar{\psi}_{2})(\psi^{j}_{2} - \bar{\psi}_{2})^{T}.
\end{align}
The sample states are then resampled and propagated to the next time step using
\eqref{eq:uupdate}, \eqref{eq:unormalize}, and \eqref{eq:upropagate}. Our
choice of sample states (sigma points) follows equation 11 of Ref.
\cite{julier2004unscented}.
Prescriptions that require fewer sigma points exist \cite{julier2002reduced}.

\section{Derivative Method \label{appendix:derivative}}
Here, we outline how to efficiently integrate the dynamics
for the derivative method using exponential integrators.
General exponential integrators break the dynamics into a linear term and a non-linear term.
For example, the dynamics for the first state derivative
are,
\begin{equation}
\frac{d}{dt} \ket{\partial_{\lambda} \psi} =
-\frac{i}{\hbar} H  \ket{\partial_{\lambda} \psi} -
\frac{i}{\hbar} (\partial_{\lambda} H) \ket{\psi}. \label{eq:gd1dyn}
\end{equation}
The linear term is $L = -\frac{i}{\hbar} H$ and the non-linear
term is $N = -\frac{i}{\hbar} (\partial_{\lambda} H) \ket{\psi}$.
With piecewise-constant controls, the exact solution to \eqref{eq:gd1dyn} is,
\begin{equation}
  \label{eq:dgeneralexp}
  \begin{aligned}
    \ket{\partial_{\lambda} \psi_{k + 1}} &= \exp(\Delta t_{k} L_{k})
    \ket{\partial_{\lambda} \psi_{k}}\\
    &+ \int_{0}^{\Delta t_{k}} \exp((\Delta t_{k} - t^{'})L_{k})
    N(t_{k} + t^{'}) dt^{'}.\\
  \end{aligned}
\end{equation}
General exponential integrators proceed by breaking the integral in \eqref{eq:dgeneralexp}
into a discrete sum, similar to the procedure
for Runge-Kutta schemes. We use a simple approximation known as the Lawson-Euler
method \cite{berland2006solving},
\begin{equation}
  \begin{aligned}
    \ket{\partial_{\lambda} \psi_{k + 1}} &\approx \exp(\Delta t_{k}L_{k})
    \ket{\partial_{\lambda} \psi_{k}}\\
    &+ \exp(\Delta t_{k}L_{k}) N_{k} \Delta t_{k}.\\
  \end{aligned}
\end{equation}
This method provides a good tradeoff between accuracy and efficiency, requiring one unique matrix
exponential computation per stage. Integration accuracy for the state derivatives
is not of the utmost importance because they are used in the robustness cost function -- as
opposed to the states themselves which are experimental parameters that must be realized
with high accuracy.

\section{Computational Performance \label{appendix:time}}
In this section we provide runtimes for our optimizations. The runtimes for
the base optimization in Sec. \ref{sec:fluxonium},
the depolarization optimization in Sec. \ref{sec:longitude},
and the robust optimizations in Sec. \ref{sec:static}
are presented in Table \ref{tab:time}
for a $Z/2$ gate at gate times which are multiples of $1/4f_{q} \sim 18$ns.
We performed optimizations on a single core of
an AMD Ryzen Threadripper 3970X 32-Core Processor @ 3.7 GHz.
Future work will parallelize the robustness methods using GPUs
\cite{leung2017speedup},
which will enable fast optimizations on high-dimensional Hilbert spaces.

\begin{table}[H]
  \centering
  \begin{tabular} {| c | c | c | c | c |}
    \hline
    \backslashbox{Method}{$t_{N} (\textrm{ns})$} & $18$ & $36$ & $72$\\
    \hline
    Base & $0.155 \pm 0.008$ & $7.0 \pm 0.4$ & $15.9 \pm 0.8$\\
    \hline
    Depol. & $1.69 \pm 0.08$ & - & -\\
    \hline
    S & $1.77 \pm 0.09$ & $48 \pm 2$ & $280 \pm 10$\\
    \hline
    U & $75 \pm 4$ & $340 \pm 20$ & $400 \pm 20$\\
    \hline
    D1 & $6.1 \pm 0.3$ & $27 \pm 1$ & $65 \pm 3$\\
    \hline
    D2 & $15.7 \pm 0.8$ & $17.3 \pm 0.9$ & $54 \pm 3$\\
    \hline
  \end{tabular}
  \caption{
    Average runtimes in seconds for $Z/2$ optimizations
    using the base, depolarization, sampling (S),
    unscented sampling (U), and the 1\textsuperscript{st}-
    and 2\textsuperscript{nd}-order derivative methods (D1, D2).
  }
  \label{tab:time}
\end{table}

\bibliography{refs}

\end{document}